\documentclass[conference]{IEEEtran}

\usepackage{epsf}
\usepackage{graphicx,color}
\usepackage{amsmath,bm}
\usepackage{amssymb}
\usepackage{epsfig,latexsym,amsmath,epsf,amssymb,amsfonts}
\usepackage{verbatim}
\usepackage{placeins}
\usepackage[hang]{subfigure}
\usepackage{epstopdf}
\usepackage{cite}
\usepackage{amsthm}
\usepackage{graphics}

\newtheorem{lemma}{Lemma}

\newcommand{\ignore}[1]{}

\usepackage{url}

\usepackage{algorithm}
\usepackage{algorithmic}

\usepackage[capitalise]{cleveref}

\usepackage{filecontents}

\usepackage{caption}
\begin{document}
%%%%%%%%%%%%%%%%%%%%%%%% CONFERENCE
%\title{Optimizing Cooperative Cognitive Radio Networks Performance with Primary QoS Provisioning}
\title{On Optimizing Cooperative Cognitive User Performance under Primary QoS Constraints}
%\author{\large Adel M. Elmahdy$^{\dag}$, xxx$^{\dag}$, yyy$^{\dag\S}$ and zzz$^*$\\ [.1in]
\author{\large Adel M. Elmahdy$^{\dag}$, Amr El-Keyi$^{\dag}$, Tamer ElBatt$^{\dag\S}$ and Karim G. Seddik$^{\star}$\\ [.1in]
\normalsize
\begin{tabular}{c}
$^{\dag}$Wireless Intelligent Networks Center (WINC), Nile University, Giza, Egypt.\\
$^{\S}$Dept. of EECE, Faculty of Engineering, Cairo University, Giza, Egypt.\\
$^{\star}$Electronics and Communications Engineering Department, American University in Cairo, New Cairo 11835, Egypt.\\
Email: adel.elmahdy@ieee.org, aelkeyi@nu.edu.eg, telbatt@ieee.org, kseddik@aucegypt.edu
\vspace{-0.1in}
\end{tabular}}
\maketitle
%%%%%%%%%%%%%%%%%%%%%%%% JOURNAL
%\title{Optimizing Cooperative Cognitive Radio Networks Performance with Primary QoS Provisioning}
%\author{\IEEEauthorblockN{Adel M. Elmahdy$^{\dag}$, Amr El-Keyi$^{\dag}$, Tamer ElBatt$^{\dag\S}$ 
%and Karim G. Seddik$^*$ \\}
%\IEEEauthorblockA{$^{\dag}$Wireless Intelligent Networks Center (WINC), Nile University, Giza, Egypt.\\
%$^{\S}$Dept. of EECE, Faculty of Engineering, Cairo University, Giza, Egypt.\\
%$^*$Electronics and Communications Engineering Department, American University in Cairo, New Cairo 11835, Egypt.\\
%{\tt adel.elmahdy@ieee.org,
%aelkeyi@nileuniversity.edu.eg, telbatt@ieee.org, kseddik@aucegypt.edu}}}
%\maketitle

%%%%%%%%%%%%%%%%%%%%%%%%%%%%%%%%%%%%%%%%%%%%%%%%%%%%%%%%%%%%%%%%%%%%%%%%%%%%%%%%%%%%%%%%%%%%%%%%%%%%%%%%%%%%%%%%%%%%%%%%%%%
\begin{abstract}
We study the problem of optimizing the performance of cognitive radio users with opportunistic real-time applications subject to primary users quality-of-service (QoS) constraints. %under non-work-conserving and work-conserving policies.
%
%In the work-conserving policy, 
Two constrained optimization problems are formulated; the first problem is maximizing the secondary user throughput while the second problem is minimizing the secondary user average delay, subject to a common constraint on the primary user average delay.
In spite of the complexity of the optimization problems, due to their non-convexity, we transform the first problem into a set of linear programs and the second problem into a set of quasiconvex optimization problems.
We prove that both problems are equivalent with identical feasible sets and optimal solutions.
%
%%Subsequently, the idea of optimizing the secondary user average delay is developed for a work-conserving policy. Due to %%the sheer complexity of such optimization problem, it is reformulated to another problem such that its solution yields an %%upper bound on the optimal secondary user delay that can be attained.
%
%%Afterwards, a practical WC-policy-based algorithm is designed in order to closely approach the optimal value of the %%secondary user delay.
%
We show, through numerical results, that the proposed cooperation policy represents the best compromise between enhancing the secondary users QoS and satisfying the primary users QoS requirements.
\end{abstract}

%\begin{IEEEkeywords}
%Cognitive radio, cooperative communications, packet delay, QoS, stable throughput region. %, numerical results.
%\end{IEEEkeywords}

%%%%%%%%%%%%%%%%%%%%%%%%%%%%%%%%%%%%%%%%%%%%%%%%%%%%%%%%%%%%%%%%%%%%%%%%%%%%%%%%%%%%%%%%%%%%%%%%%%%%%%%%%%%%%%%%%%%%%%%%%%%
\section{Introduction}
%\subsection{Literature Review}
%\vspace{-10pt}
\makeatletter{\renewcommand*{\@makefnmark}{}
\footnotetext{\hrule \vspace{0.05in}
This work was made possible by grants number NPRP 4-1034-2-385 and
NPRP 5-782-2-322 from the Qatar National Research Fund (a member of
Qatar Foundation). The statements made herein are solely the responsibility
of the authors.
%
%Part of this work has been submitted in preliminary form to Asilomar
%Conference on Signals, Systems and Computers, Pacific Grove, CA, 2015.
}\makeatother}
The concept of cognitive radio was stimulated by the problem of severe underutilization of the licensed spectrum, in addition to the spectrum scarcity \cite{survey_Liang}.
The cognitive radio technology aims at exploiting the spectrum holes and, hence, efficiently utilizing the precious wireless spectrum.
Cognitive radio networks consist of licensed primary users (PUs), who may not transmit data the whole time, and unlicensed secondary users (SUs) having sensing capability of the spectrum in order to detect and exploit the spectrum holes for the transmission of their packets.
The coexistence of SUs with PUs is subject to the condition that a certain level of QoS is guaranteed to the PUs.

The notion of cooperative wireless communications hinges on the broadcast nature of the wireless medium.
A data transmission between a source and a destination might be received and decoded by intermediate nodes that could act as ``relays" \cite{Tse,Kramer}.
One of the advantages of cooperative communications is improving the performance of wireless networks since the relay can retransmit packets that are not successfully decoded by the destination.
%Consequently, the relay can assist the source in delivering the lost data over the direct link.
%
%From a different perspective, cooperative communications can be thought of as an analogous way to using multiple antennas in order to achieve spatial diversity \cite{CoopComm_Foschini,CoopComm_Telatar}.
%
A significant part of the literature was dedicated to studying the idea of cooperative communications from the perspective of the physical (PHY) layer, e.g., \cite{Gamal,CoopComm_Sadek_decod_forwrd}.
On the other hand, leveraging cooperative communications within the context of cognitive radio networks has promised considerable performance gains and has been studied in the literature at the medium access control (MAC) layer. 
Recently, there has been growing interest in cognitive relaying networks where the SU can assist the PU in delivering its packets to the destination, e.g.,~\cite{CoopCRN_Simeone_Relay,Sadek,Kompella_stableThroughput,Ashour_Journal,Adel_Conf}.
Such cooperation would be beneficial to both the PU and the SU. The PU will reliably transmit its packet to the destination through the SU when the data is lost over the direct link.
Therefore, the number of time slots, in which the SU can access the channel and transmit its own packets, will increase.
For example, a cognitive interference channel is studied in \cite{CoopCRN_Simeone_Relay} where the SU acts as a relay for the PU traffic. A power allocation scheme at the SU is designed in order to maximize the stable throughput of the secondary link for a fixed throughput of the primary link. It should be noted that the optimization problem does not impose any constraints on the average packet delay encountered by the PU.
In \cite{Sadek}, two cooperative cognitive multiple-access protocols are proposed in a network that consists of $M$ source terminals, a relay node, and a common destination node. The performance gains of the proposed protocols over conventional relaying strategies are demonstrated in terms of the maximum stable throughput region and the delay performance.
In \cite{Kompella_stableThroughput}, the stable throughput region is characterized for a cooperative cognitive network with a fixed scheduling probability. Specifically, the secondary link is allowed to share the channel along with the primary link, and the secondary node cooperatively relays those packets that it decodes successfully from the primary node, but are not decoded by the primary destination.

%\subsection{Motivation and Related Work}
The motivation of this paper is the emergence of opportunistic real-time (ORT) traffic in cognitive radio networks, in general, and cooperative cognitive radio networks in particular.
This arises in operating regimes where the PUs and SUs are using multimedia applications, demanding high throughput and stringent delay requirements.
%
%%%%%Providing a better QoS for real-time applications of secondary users can be 
%%%%%an economic incentive to motivate the primary users to share their idle
%%%%%spectrum with the secondary users. However, this is on condition that 
%%%%%the QoS of the primary users is not adversely affected.
% There are many techniques on methods for the coexistence of PUs and SUs in cognitive radio networks as long as the SUs do not degrade the performance of the PUs \cite{survey_Liang}.
%
An example from the cooperative communications perspective is a protocol-level cooperation, proposed in \cite{R_CoopAcc}, for a wireless multiple-access system with probabilistic transmission success. The system comprises $N$ users and a common destination. Each user is considered as a source and, at the same time, a potential relay.
For the two-user case, user~$1$ has one queue for its own packets while user~$2$ has two queues; one for its own packets and the other for the relayed packets from user~$1$.
The proposed cooperation policy always gives user~$1$ higher priority than user~$2$ to access the channel and transmit its packets.
The shortcoming of applying this policy in the framework of cooperative cognitive radio networks is giving strict priority to user 1 (PU) packets, possibly yielding average PU packet delay much stricter than it can tolerate (i.e., over designing the system for the PU).
In other words, according to this policy, user 2 (or SU) packets may experience severe delay while PU packets can tolerate higher delays (e.g., delay-insensitive traffic).

Another example from the perspective of cognitive relaying is a non-work-conserving (non-WC) cooperation policy, proposed in \cite{Ashour_Journal}, for a cooperative cognitive radio network with two tunable parameters.
The first one is 
the probabilistic relaying parameter
(i.e., probabilistic admission control of the PU packets in the SU relaying queue), 
and the second one is 
the randomized service parameter at the SU (i.e., probabilistic selection between two queues; one for the SU packets and the other for the relayed PU packets).
%The authors thoroughly characterize the stable throughput region and the average packet delay of the network.
The fundamental delay-throughput tradeoff is studied and two optimization problem are formulated;
the first problem is to minimize the average packet delay encountered by the PU subject to queues stability constraints. The second problem is to minimize the average packet delay encountered by the SU subject to the same constraints.
It should be noted that the problem of optimizing the SU performance does not take into consideration the the average delay of the PU. Thus, the optimal values of the randomized service and probabilistic relaying parameters result in severe delay for the PU packets. It is needless to mention that the essence of cognitive radio networks is maintaining a certain level of QoS for the PU while serving the SU.
In \cite{Adel_Conf}, the authors characterize the stable throughput region of the system in \cite{Ashour_Journal} when the relaying queue at the SU has limited capacity. In addition, the packet admission and queue selection probabilities are
dependent on the relaying queue length at each time slot.

In summary, the portion of time at which the PU uses delay-insensitive traffic (e.g., web browsing), the PU can tolerate some delay to provide better quality to the SU delay-sensitive traffic (e.g., video streaming) in exchange for some incentives.
Therefore, the network resources are efficiently utilized and the cognitive user performance is optimized with primary user QoS provisioning.

%\subsection{Summary of Results}
Unlike \cite{R_CoopAcc,Ashour_Journal,Adel_Conf}, this paper optimizes the QoS of real-time applications of SUs while preserving the QoS of PUs in cooperative cognitive radio networks. 
%%for both non-WC as well as work-conserving (WC) cooperation policies.
%Perhaps the closest to our work is the cooperation scheme proposed by 
%\cite{Ashour_Conf}, coined ``unconstrained partial cooperation'', in which 
%an optimization problem is formulated to minimize the SU average
%delay subject only to queues stability constraints for
%a non-work-conserving (non-WC) system.
%It does not take into consideration the average delay of the PU and, hence,
%the PU packets suffer from severe delay.
%It is needless to mention that the essence of cognitive radio networks
%is maintaining a certain level of QoS for the PU while serving the SU.
%However, unlike \cite{Ashour_Conf}, this work investigates the
%possibility of improving the QoS of real-time applications
%for the SU while preserving the QoS of the PU in cooperative
%cognitive radio networks for both non-WC as well as
%work-conserving (WC) cooperation policies.
%
%Essentially, we extend the work conducted in \cite{Ashour_Conf}.
%
The main contributions of this paper are as follows.
%In the first part of the paper, 
%%We investigate a cooperative cognitive radio network that operates under a non-WC cooperation policy.
%%More specifically, 
We formulate two distinct optimization problems.
The first problem maximizes the SU throughput while the second problem minimizes the average packet delay encountered by the SU. Both problems are subject to a constraint on the average delay encountered by the PU packets.
%Furthermore, the QoS of the PU traffic should
%not be affected in cognitive radio networks.
%Hence, our aim in this paper is to investigate
%such tradeoff and find out the best cooperation strategies that
%solve the proposed optimization problems.
%%
Although the formulation of each problem yields a non-convex optimization problem, the first problem is transformed into a set of linear programs, whereas the second problem is transformed into a set quasiconvex optimization problems. 
In addition, we prove that the problem of optimizing the SU throughput is equivalent to optimizing the SU delay for the studied system and they have the same feasible solution set.
The numerical results show that the proposed optimal %partial --> optimal in all 
%%non-WC 
cooperation policy guarantees that the throughput and the average packet delay of the SU are enhanced, while honoring the average PU packet delay constraint.
%Finally, the simulation results reveal a partial cooperation policy
%%for each optimization problem
%such that the throughput and the
%average packet delay of the SU are improved, and, at the same time,
%the average packet delay of the PU is not degraded.

%%In the second part of the paper, we study the problem of optimizing the SU packet delay subject to a constraint on the PU %%packet delay for a WC cooperation policy towards more efficient resource utilization.
%%It is worth mentioning that the cooperation policy considered in the first part of the paper leads to a non-WC policy %%because it is susceptible to wasting idle time slots. 
%%However, the mathematical analysis of its average packet delay is mathematically tractable. 
%%On the other hand, the derivation of closed-form expressions for average packet delay of the WC policy is complex because %%the analysis involves the interaction of three dependent queues \cite{S_2IntQ}. 
%%In order to alleviate these hurdles, the target optimization problem is reformulated to another problem whose solution is %%an upper bound on the optimal SU delay. 
%%Afterwards, the numerical results show that the proposed suboptimal WC policy outperforms the non-WC policy studied in %%the first part of the paper. Finally, a practical WC-policy-based algorithm is proposed in order to closely approach the %%optimal solution of the target optimization problem.

%W%\subsection{Organization}
The rest of this paper is organized as follows.
The system model is given in Section~II.
The SU throughput and the average SU packet delay are characterized in Section~III.
The problem of optimizing the SU performance subject to a constraint on the PU delay is 
formulated and solved in Section~IV.
%%for a non-WC cooperation policy.
%The problem of optimizing the SU throughput subject to a constraint on the PU delay is formulated and solved in Section~IV, whereas the problem of optimizing the SU delay subject to a constraint on the PU delay is formulated and solved in Section~V for a non-WC cooperation policy.
%%In Section~V, the problem of minimizing the SU delay subject to the same constraint for a WC cooperation policy is %%investigated and a WC-policy-based algorithm is proposed in order to closely approach the optimal SU delay.
The numerical results are presented and discussed in Section~V.
Finally, the paper is concluded in Section~VI.

%%%%%%%%%%%%%%%%%%%%%%%%%%%%%%%%%%%%%%%%%%%%%%%%%%%%%%%%%%%%%%%%%%%%%%%%%%%%%%%%%%%%%%%%%%%%%%%%%%%%%%%%%%%%%%%%%%%%%%%%%%%
\section{System Model}
\begin{figure}
\centering
\includegraphics[width=0.9\linewidth]{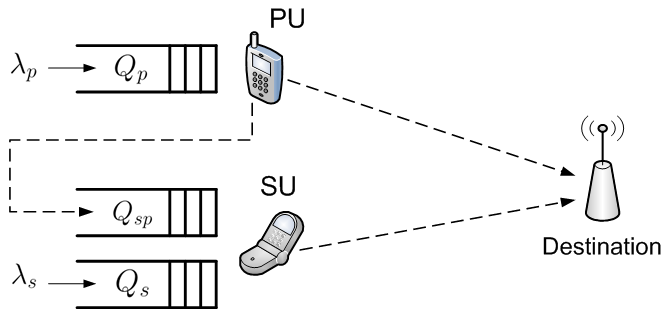}
\caption{The system model.
The dashed lines represent communication links between nodes.}
\label{fig:SysModel}
\vspace{-3mm}
\end{figure}

We consider the cooperative cognitive radio network depicted in Fig.~\ref{fig:SysModel}.
The network comprises two users (e.g., two mobile stations), a PU and a SU, 
and a common destination (e.g., a base station).
The PU is equipped with a queue, $Q_p$, for the primary user packets.
On the contrary, the SU has two queues, $Q_{s}$
and $Q_{sp}$. $Q_{s}$ is intended for the secondary user packets,
whereas $Q_{sp}$ is intended for the packets that are overheard,
decoded and enqueued from the PU. All queues are assumed to be of
infinite length.
The assumption of infinite queue length is reasonable when the queue size 
is much larger than the packet size.

We assume a time-slotted system where the transmission of a packet takes exactly one time slot.
%We assume a time-slotted system where a packet fits exactly in one time slot.
The packet arrival processes at $Q_p$ and $Q_s$ are modelled as
Bernoulli random processes with rates $\lambda_p$ and $\lambda_s$
packets per time slot, respectively, where $0 \!\leq\! \lambda_p \!\leq\! 1$ and $0 \!\leq\! \lambda_s \!\leq\! 1$. 
The packet arrival processes are assumed independent
from each other and packet arrivals at each queue are independent
and identically distributed (i.i.d.) across time slots.
The evolution of the length of the $j^{th}$ queue
is characterized as
\begin{equation} \label{eq:queue_evolution}
Q_j^{t+1} = \left(Q_j^t - Y_j^t\right)^+ + X_j^t, \text{ for } j \in \{p, sp, s\},
\end{equation}
where $\left(\centerdot\right)^+ = \max(\centerdot,0)$. $Q_j^{t}$ denotes the number of packets
of the $j^{th}$ queue at time slot $t$. $X_j^t$ and $Y_j^t$ are
binary random variables that represent the number of packets
that arrive at (or depart from) the $j^{th}$ queue at time slot $t$,
respectively.
A~positive acknowledgment packet (ACK) is sent by a receiving node that successfully
decodes a packet, and it is heard by all other nodes in the network.
In the cooperative cognitive radio network illustrated in Fig.~\ref{fig:SysModel}, 
an ACK is sent from either the destination or the SU.
The length of an ACK is assumed to be very short compared to the slot duration.
It is also assumed that the errors as well as the delay in the acknowledgment feedback channel are negligible.
This assumption is justified by employing low rate codes in the feedback channel.

The prime causes of the degradation of wireless link quality are multipath fading, additive noise, and signal attenuation.
We assume that the random processes modeling the channel gains and noise are stationary.
The probability of wireless link outage is the probability that the transmission rate of a source exceeds the instantaneous link capacity. For fixed-rate transmission over the primary and secondary links, the link outage probability is inversely proportional to the average signal-to-noise ratio (SNR) at the receiver. Therefore, the link outage occurs when the average SNR is below the threshold at which the receiver can decode the incoming packets without errors.
Throughout this paper, the quality of wireless links is abstracted by the likelihood that a node correctly decodes a packet.
The probability of successful packet reception, i.e., the probability of no link outage, between the PU and the destination, the SU and the destination, and the PU and the SU are denoted by $h_{pd}$, $h_{sd}$ and $h_{ps}$, respectively.
%In this paper, the quality of the wireless channel is abstracted
%by the likelihood that a node correctly decodes a packet. The
%prime causes of link quality degradation
%are signal attenuation, noise and multipath fading.
%We assume that the random processes modeling the channel impairments
%are stationary.
%Channel outage occurs when the average receive signal-to-noise ratio
%is below a threshold at which the destination can decode the 
%received packets without error.
%The probability of successful packet reception, 
%i.e., the probability of no link outage,
%between the PU and the destination, the SU and
%the destination, and the PU and the SU are denoted by
%$h_{pd}$, $h_{sd}$ and $h_{ps}$, respectively.

Similar to \cite{Sadek,Ashour_Journal,Adel_Conf}, the SU is assumed to perfectly know the state of the PU of whether it is backlogged or idle and, hence, there is no interference in our system.
%One approach to accomplish this objective is through the transmission of one bit of information from the PU to the SU via the feedback channel. 
A possible approach to accomplish this objective is via sensing the communication channel by the SU in order to detect the time slots at which the PU is idle. This can be achieved by using detectors that have high detection probability at the SU. If the SU causes interference to the PU due to a misdetection, the interference structure could be leveraged in the detection process. Nevertheless, this is beyond the scope of this paper.
%Similar to \cite{Ashour_Conf,Adel_Conf}, perfect sensing process is assumed at the secondary
%transmitter and, therefore, there is no interference between the PU
%and the SU in the studied system.
%The assumption is convenient for the analytical formulation of the target problem.
%

We adopt the following cooperation policy at the MAC layer.
\subsection{When the PU is backlogged}
$Q_p$ immediately transmits the head-of-line (HOL) packet to the destination since it is the spectrum owner. 
Three potential cases arise:
\begin{itemize}
\item 
If the packet is successfully decoded by the destination, an ACK is broadcast and the packet is dropped from the system, regardless of whether the SU successfully decodes it or not.
\item 
If the packet is successfully decoded by the SU, but is not decoded by the destination, the packet is stored in $Q_{sp}$ with probability $a$.
If admitted, the SU broadcasts an ACK and the packet is dropped from $Q_p$.
\item 
If neither the SU nor the destination decodes the packet, then it is kept in $Q_p$ for future retransmission.
\end{itemize}
\subsection{When the PU is idle}
\begin{itemize}
\item 
The channel is accessed by the SU and a packet is transmitted either from $Q_s$ with probability $b$ or from $Q_{sp}$ with probability $1-b$.
\item 
If the destination successfully decodes the packet, an ACK is broadcast and the packet is dropped from the system. 
Otherwise, the packet is kept in its respective queue for future retransmission.
\end{itemize}
It is worth noting that the aforementioned cooperation policy is non-WC \cite{Wolff}.
The reason lies behind the possibility that the system might have packets in its queues, yet the slot is wasted.
A typical case occurs when the SU accesses the channel and an empty queue is selected for transmission while the other queue is non-empty.
It should be noted that we focus in this paper on the aforementioned non-WC cooperation policy, due to its mathematical tractability \cite{Ashour_Journal}.
%%However, we relax this assumption and study the WC cooperation policy in Section~V despite its reported complexity %%\cite{S_2IntQ} attributed to the interaction of three dependant queues.
%The delay analysis of the WC version is notoriously complicated since the analysis hinges on the moment generating function of the joint lengths of three dependant queues \cite{S_2IntQ}.

%%%%%%%%%%%%%%%%%%%%%%%%%%%%%%%%%%%%%%%%%%%%%%%%%%%%%%%%%%%%%%%%%%%%%%%%%%%%%%%%%%%%%%%%%%%%%%%%%%%%%%%%%%%%%%%%%%%%%%%%%%%
\section{Background: Throughput and Average Delay Characterization}
In this section, we characterize the service rates for various queues as well as the arrival rate for the relay queue at the SU with the aid of different probabilistic events for the cooperative cognitive radio network depicted in Fig.~\ref{fig:SysModel}. Next, the stability of the queues of the network is established.
Finally, the expressions for average packet delay of the PU and SU are~presented.

A packet departs $Q_p$ in two cases:
1) if it is decoded by the destination (i.e., the direct link is not in outage), 
or 2) if it is not decoded by the
destination, yet, is decoded by the SU and admitted by $Q_{sp}$.
Thus, the service rate of $Q_p$, $\mu_p$, is given~by
\begin{equation} \label{eq:mu_p2}
\displaystyle\mu_p = h_{pd} + \left(1-h_{pd}\right) h_{ps} a.
\end{equation}
On the other hand, a packet in $Q_s$ is served when $Q_p$ is empty, 
$Q_s$ is selected for transmission, and there is no channel outage
between the SU and the destination. Therefore,
the service rate of $Q_s$, $\mu_s$, is given by 
%the service rate of $Q_s$, $\mu_s$, for a given arrival 
%rate of the PU packets, $\lambda_p$, is characterized as
\begin{equation} \label{BKG_mu_s}
\displaystyle\mu_s = b h_{sd} \left( 1 - \frac{\lambda_p}{\mu_p} \right).
\end{equation}
Similarly, the service rate of $Q_{sp}$, $\mu_{sp}$, is given by
\begin{equation} \label{eq:mu_sp}
\displaystyle\mu_{sp} = \left(1-b\right) h_{sd} \left(1-\frac{\lambda_p}{\mu_p}\right).
\end{equation}
Furthermore, a packet is buffered at $Q_{sp}$ when $Q_p$ is non-empty,
the direct link is in outage but the link between the PU and SU is not in outage, 
and the packet is admitted by $Q_{sp}$. 
Consequently, the packet arrival rate to $Q_{sp}$, $\lambda_{sp}$, is defined~as
\begin{equation} \label{eq:lmbda_sp}
\displaystyle\lambda_{sp} = a \left(1-h_{pd}\right) h_{ps} 
\frac{\lambda_p}{\mu_p}.
\end{equation}

%A packet in $Q_s$ is served when $Q_p$ is empty, 
%$Q_s$ is selected for transmission, and there is no channel outage
%between the SU and the destination. Therefore,
%the service rate of the SU packets for a given arrival 
%rate of the PU packets is defined as 
%\begin{equation}
%\displaystyle\mu_s = b h_{sd} \left( 1 - \frac{\lambda_p}{\mu_p} \right),
%\end{equation}
%where the PU service rate is given by
%\begin{equation} \label{eq:mu_p2}
%\mu_p = h_{pd} + \left(1-h_{pd}\right) h_{ps} a.
%\end{equation}
%Note that a packet departs $Q_p$ in two cases:
%1) if it is decoded by the destination (i.e., the direct link is not in outage), 
%or 2) if it is not decoded by the
%destination, yet, decoded by the SU and admitted by $Q_{sp}$. 
%Also note that the stability of $Q_s$ is guaranteed when the 
%following condition is satisfied
%\begin{equation}
%\displaystyle \lambda_s < \mu_s.
%\end{equation}

The stability of a queue is characterized by Loynes' theorem~\cite{L_QStab}.
When the arrival and service processes of a queue are stationary,
the queue is stable if and only if the packet arrival rate is strictly less than the packet service rate. Otherwise, the queue is unstable.
Accordingly, the stability of the queues for the studied network is characterized by the following~inequalities
\begin{equation} \label{eq:stability}
\displaystyle \lambda_p < \mu_p, \quad \lambda_s < \mu_s, \quad \lambda_{sp} < \mu_{sp}.
\end{equation}

The average delay experienced by the PU packets 
and the SU packets can be characterized by applying Little's law \cite{Kleinrock} as follows
\begin{equation} \label{eq:D_p and D_s}
\displaystyle D_p = \frac{N_{p}+N_{sp}}{\lambda_p}, \quad\:
\displaystyle D_s = \frac{N_{s}}{\lambda_s},
\end{equation}
%%%\begin{equation} \label{eq:D_p}
%%%\displaystyle D_p = \frac{N_{p}+N_{sp}}{\lambda_p},
%%%\end{equation}
%%%\begin{equation} \label{eq:D_s}
%%%\displaystyle D_s = \frac{N_{s}}{\lambda_s},
%%%\end{equation}
where $N_{p}$, $N_{sp}$ and $N_{s}$ are the average queue 
length of $Q_{p}$, $Q_{sp}$ and $Q_{s}$, receptively.
$N_{p}$ is obtained by direct application
of the Pollaczek-Khinchine formula \cite{Kleinrock} on $Q_p$, a
discrete-time M/M/1 queue with Bernoulli arrival rate $\lambda_p$
and geometrically distributed service rate $\mu_p$; it is given by
\begin{equation} \label{eq:N_p}
\displaystyle N_p = \frac{\lambda_p-\lambda_p^2}{\mu_p - \lambda_p}.
\end{equation}
On the other hand, the expressions for $N_{sp}$ and $N_{s}$, 
in terms of $\mu_p$, are given by 
\begin{equation} \label{eq:N_sp2}
\displaystyle N_{sp} = 
\frac{\lambda_p \!\left(\mu_p \!\!-\!\! h_{pd}\right)
\!\left(\overline{b} h_{sd} \overline{\mu_p} \lambda_p 
\!\!-\!\! \left(\mu_p \!\!-\!\! h_{pd}\right)\! \mu_p \lambda_p
\!\!-\!\! h_{pd} \lambda_p \!\!+\!\! \mu_p^2\right)\!}
{\mu_p \!\left(\mu_p \!\!-\!\! \lambda_p\right)
\!\left(\overline{b} h_{sd} \!\left(\mu_p \!\!-\!\! \lambda_p\right) 
\!-\! \lambda_p \!\left(\mu_p \!\!-\!\! h_{pd}\right)
\right)\!},
\end{equation}
\begin{equation} \label{eq:N_s2}
\displaystyle N_{s} = 
\frac{\left(\mu_p \!\!-\!\! \lambda_p\right)
\!\left(b h_{sd} \!\left(\mu_p \!\!-\!\! \lambda_p\right) \!-\! \lambda_s \mu_p\right)}
{b h_{sd} \lambda_p \lambda_s \overline{\mu_p} 
\!+\! \left(\lambda_s \!\!-\!\! \lambda_s^2\right)\!
\left(\mu_p \!\!-\!\! \lambda_p\right)\! \mu_p},
\end{equation}
%\begin{equation} \label{eq:N_sp2}
%\displaystyle N_{sp} = \frac{m \lambda_p^2 + n \lambda_p}{\alpha \lambda_p^2 + \beta \lambda_p + \gamma}
%\end{equation}
%where
%\begin{IEEEeqnarray}{lCl} \label{eq:Nsp_paramt2}
%  \displaystyle
%  m & = &
%  \left(\mu_p \!-\! h_{pd}\right) \! \biggl(\! \frac{\left(1 \!-\! b\right)h_{sd} \! - \! h_{pd}}{\mu_p}
%  - \left(1 \!-\! b\right) \! h_{sd} - \left(\mu_p \!-\! h_{pd}\right) \! \biggr)
%  \nonumber\\
%  n & = & \left(\mu_p - h_{pd}\right) \mu_p
%  \nonumber\\
%  \alpha & = & \left(1 - b\right) h_{sd} + \left(\mu_p - h_{pd}\right)
%  \nonumber\\
%  \beta & = & \mu_p \left(-2 \left(1 - b\right) h_{sd} - \left(\mu_p - h_{pd}\right)\right)
%  \nonumber\\
%  \gamma & = & \left(1 - b\right) h_{sd} \mu_p^2
%\end{IEEEeqnarray}
%
%\begin{equation} \label{eq:N_s2}
%\displaystyle N_{s} = \frac{\lambda_p \lambda_s A + \left(\lambda_s^2 - \lambda_s\right) B \left(B+\lambda_p\right)}
%{B C}
%\end{equation}
%where
%\begin{IEEEeqnarray}{lCl} \label{eq:Ns_paramt2}
%  \displaystyle
%  A & = & b h_{sd} \left(\mu_p - 1\right) \nonumber\\
%  B & = & \mu_p - \lambda_p \nonumber\\
%  C & = & \left(\lambda_s - b h_{sd}\right) \mu_p + b h_{sd} \lambda_p.
%\end{IEEEeqnarray}
where $\overline{b} = 1-b$ and $\overline{\mu_p} = 1-\mu_p$.
The proof of these expressions directly follows the approach in \cite{S_2IntQ}. %\cite[Theorem~5]{Ashour_Journal}
In particular, $N_{sp}$ and $N_{s}$ are evaluated by applying the moment generating function approach
and analyzing the interaction of the joint lengths of the dependent queues 
$Q_{p}$~and~$Q_{sp}$, and $Q_{p}$~and~$Q_{s}$, respectively~\cite{Ashour_Journal}.
%The proof is eliminated here for space limitation.

%%%%%%%%%%%%%%%%%%%%%%%%%%%%%%%%%%%%%%%%%%%%%%%%%%%%%%%%%%%%%%%%%%%%%%%%%%%%%%%%%%%%%%%%%%%%%%%%%%%%%%%%%%%%%%%%%%%%%%%%%%%
\section{Optimizing the Secondary User Performance with Primary QoS Provisioning}
In this section, the problem of optimizing the SU QoS under the PU QoS constraints is formulated and solved. In the first part, the SU throughput is optimized subject to a constraint on the average PU delay. In the second part, the average SU delay is optimized subject to the same constraint.
\subsection{Optimizing the Secondary User Throughput}
In this subsection, we investigate the problem of maximizing the SU throughput subject to a constraint on the average PU packet delay, $D_{p}$. 
%for a given arrival rate of the PU packets, $\lambda_p$,
%
It is worth mentioning that introducing a constraint on $D_p$ is stricter than the stability constraint and, hence, implies the stability of $Q_p$, i.e., the queue length is guaranteed not to grow to infinity.
Therefore, there is no need for a $Q_p$ stability constraint in the sought formulation.
%Therefore, the stability condition on $Q_p$ will be omitted from the constraints of the optimization problem.
%
Consequently, the target constrained optimization problem is formulated as
\begin{eqnarray} \label{min_1}
    \displaystyle \textbf{P1: } \quad \max_{a, b} \!\!&&\!\!  \displaystyle b \: h_{sd} \left(1 -
        \frac{\lambda_p}{\mu_p}\right) \nonumber\\
    \text{s.t. } \!\!&&\!\! 0 \leq a \leq 1 \nonumber\\
    \!\!&&\!\! 0 \leq b \leq 1 \nonumber\\
    \!\!&& \!\! \displaystyle \mu_p = h_{pd} + \left(1-h_{pd}\right) h_{ps} a \nonumber\\
    \!\!&& \!\! \displaystyle \frac{N_p+N_{sp}}{\lambda_p} \leq \psi,
\end{eqnarray}
where the objective function is simply the SU packet service rate, $\mu_s$. $N_p$ and $N_{sp}$ are given by \eqref{eq:N_p} and \eqref{eq:N_sp2}, respectively,
whereas $\psi$ specifies the maximum average packet delay that the PU can tolerate. 
In real systems, the delay sensitivity of the PU applications should map to the value of $\psi$ accordingly.
\textbf{P1} is non-convex since the Hessian of the objective function is not negative semidefinite.
Our goal is to convert \textbf{P1} to a set of linear programs
%, in which the objective and constraint functions are all affine \cite{ConvxOpt}, and 
that can be solved for the optimal in an iterative manner as shown next.

Towards this objective, we go through a number of steps.
First, the range of possible values of $\mu_p$ is defined as
\begin{equation} \label{eq:mu_p_ineq}
h_{pd} \leq \mu_p \leq h_{pd}+\left(1-h_{pd}\right)h_{ps}.
\end{equation}
%%%\begin{equation} \label{eq:mu_p_ineq}
%%%\max\left(\frac{\lambda_p}{\rho},h_{pd}\right) \leq \mu_p \leq h_{pd}+\left(1-h_{pd}\right)h_{ps},
%%%\end{equation}
%%%where
%%%\begin{equation} \label{eq:rho}
%%%\rho = \frac{h_{sd} \left(1-b\right)}{h_{sd} \left(1-b\right) + a h_{ps} \left(1-h_{pd}\right)}.
%%%\end{equation}
%
This inequality can be readily verified from \eqref{eq:mu_p2}.
The service rate of the PU packets, $\mu_p$,
depends on the packet admission probability,~$a$.
Since $\:0 \!\leq\! a \!\leq\! 1$, we can accordingly specify the lower and upper bounds on $\mu_p$ as shown~in~\eqref{eq:mu_p_ineq}.
%%%\begin{equation} \label{eq:app1}
%%%h_{pd} \leq \mu_p \leq h_{pd} + h_{ps} \left(1 - h_{pd}\right).
%%%\end{equation}
%
%%%Moreover, from \eqref{eq:mu_sp}, \eqref{eq:lmbda_sp} and \eqref{eq:stability}, we get 
%%%%Moreover, we apply Loynes' theorem \cite{L_QStab} to  establish the stability of $Q_{sp}$ as follows
%%%\begin{equation}
%%%\displaystyle a \left(1-h_{pd}\right) h_{ps} \frac{\lambda_p}{\mu_p}
%%%< \left(1 - b\right) h_{sd} \left(1 - \frac{\lambda_p}{\mu_p}\right).
%%%\end{equation}
%%%Rearranging the terms yields
%%%\begin{equation} \label{eq:app2}
%%%\lambda_p < \rho \mu_p,
%%%\end{equation}
%%%where $\rho$ is given by \eqref{eq:rho}. Combining \eqref{eq:app1} and
%%%\eqref{eq:app2}, we get~\eqref{eq:mu_p_ineq}.
%
Second, we fix $\mu_p$ and then run \textbf{P1} iteratively for every possible value of $\mu_p$. 
Therefore, the only variable in the reformulated optimization problem is $b$, 
while $\mu_p$, and consequently $a$, would
be constant in each iteration where the optimization problem is solved.
It is evident from \eqref{BKG_mu_s} that $\mu_s$~is an affine function
in $b$. %and, hence it is quasiconvex in $b$.
On the other hand, it can be shown through \eqref{eq:N_p} and \eqref{eq:N_sp2}
%and substitutions from \eqref{eq:Nsp_paramt2} 
that $D_p$ is a quasiconvex
function in $b$~since it is a convex over concave function.
As a result, the constraint on the delay encountered by the PU packets is
the $\psi$-sublevel set of the quasiconvex function $D_p$, which can
be represented as the $0$-sublevel set of the convex function~$\phi_{\psi}$
that is given~by
\begin{IEEEeqnarray}{lcl} \label{eq:phi}
\displaystyle \phi_{\psi} & \:=\: &
\lambda_p \!\left(\mu_p \!\!-\!\! h_{pd}\right)
\!\left(\overline{b} h_{sd} \overline{\mu_p} \lambda_p 
\!\!-\!\! \left(\mu_p \!\!-\!\! h_{pd}\right)\! \mu_p \lambda_p
\!\!-\!\! h_{pd} \lambda_p \!\!+\!\! \mu_p^2\right)\!
\nonumber\\
& & \:-\: \mu_p \!\left(\mu_p \!\!-\!\! \lambda_p\right)
\!\left(\overline{b} h_{sd} \!\left(\mu_p \!\!-\!\! \lambda_p\right) 
\!-\! \lambda_p \!\left(\mu_p \!\!-\!\! h_{pd}\right)\right)\!
\left(\lambda_p \psi \!\!-\!\! N_p\right),
\nonumber\\
\end{IEEEeqnarray}
%\begin{equation} \label{eq:phi}
%\phi_{\psi} = \left(m \lambda_p^2 + n \lambda_p\right) -
%\left(\alpha \lambda_p^2 + \beta \lambda + \gamma\right) \left(\lambda_p \psi - N_p\right)
%\end{equation}
where $N_{p}$ is given by \eqref{eq:N_p}.
%and $m$, $n$, $\alpha$, $\beta$, $\gamma$ are given by \eqref{eq:Nsp_paramt2}.
Note that $\phi_{\psi}$ is an affine function of $b$.
%Note that $m$, $\alpha$, $\beta$, $\gamma$ are affine functions of $b$, and
%$n$, $N_{p}$ are constants.
%
Thus, \textbf{P1} can be cast as the following optimization problem
\begin{algorithmic}
\FOR{$\displaystyle\mu_p = h_{pd} : \delta : h_{pd} + \left(1 - h_{pd}\right)h_{ps}$}
\vspace{-3mm}
\STATE{
    \begin{eqnarray} \label{min_2}
        \displaystyle \textbf{P2: } \quad g_{2}\!\left(\mu_p\right)=\max_{b} \!\!&&\!\! b \: h_{sd} \left(1 -
        \frac{\lambda_p}{\mu_p}\right) \nonumber\\
        \text{s.t. } \!\!&&\!\! 0 \leq  b \leq 1 \nonumber\\
        \!\!&& \!\! \displaystyle \phi_{\psi} \leq 0
    \end{eqnarray}
} \ENDFOR \RETURN  $\displaystyle\max_{\mu_p} g_{2}\!\left(\mu_p\right)$,
\end{algorithmic}
where $\delta$ is a pre-specified increment value for $\mu_p$.
\begin{figure*}[!t]
\normalsize
\begin{equation}%\small %%KG
\label{two_Col_eqn} %omit{\displaystyle}
b^*\!\left(\mu_p\right) =
\min \left(1,\: 1 - \displaystyle \frac
{
\lambda_p^2 \left(\mu_p \!-\! h_{pd}\right) \! \left(\!\frac{-h_{pd}}{\mu_p}-\left(\mu_p \!-\! h_{pd}\right)\!\right)
+ \lambda_p \mu_p \left(\mu_p \!-\! h_{pd}\right)
- \left(\lambda_p\psi \!-\! N_p\right) \! \left(\mu_p \!-\! h_{pd}\right) \! \left(\lambda_p^2 \!-\! \lambda_p\mu_p\right)
}
{
-h_{sd} \left(\frac{\lambda_p^2}{\mu_p} \left(\mu_p \!-\! h_{pd}\right) \! \left(1 \!-\! \mu_p\right)
- \left(\lambda_p\psi \!-\! N_p\right) \! \left(\lambda_p^2 \!-\! 2\lambda_p\mu_p \!+\! \mu_p^2\right)\!\right)
}\right).
\end{equation}
\hrulefill
\vspace{-0.03in} %%KG(-0.2in)
\end{figure*}
We can see that this optimization problem is a low complexity line search in the interval 
$\left[\:\!h_{pd},\: h_{pd} \!+\! \left(1\!-\!h_{pd}\right)h_{ps}\:\!\right]$.
%It should be noted that the objective function of \textbf{P2}, $\mu_s$, is linear in $b$. 
%Therefore,
Since the objective and constraint functions of \textbf{P2} are all affine,
\textbf{P2} is a linear program for each iteration on~$\mu_p$~\cite{ConvxOpt}. %, with a variable $b$. %%AE2
Note that, for a given~$\mu_p$,  we can calculate $a$ from \eqref{eq:mu_p2}.
%A closed-form expression of the solution of \textbf{P2} can be shown to
%be \eqref{two_Col_eqn}, on the next page, via simple algebraic manipulations.
A closed-form expression of the solution of \textbf{P2} is characterized by the following lemma.
\begin{lemma}
\label{lemma0}
For a given $\mu_p$, the optimal solution of \textbf{P2}, $b^*\!\left(\mu_p\right)$, is given by \eqref{two_Col_eqn} (shown on the next page).
\end{lemma}
\begin{IEEEproof}
\label{proof0}
It is evident that the objective function of \textbf{P2} monotonically increases with $b$, where $0 \!\leq\! b \!\leq\! 1$. However, $b^*\!\left(\mu_p\right)$ must satisfy the constraint $\phi_{\psi} \leq 0$, which can be rewritten, via simple algebraic manipulations, as $b \leq f(\mu_p, \psi)$, where $f(\mu_p, \psi)$ is the second term of the $\min$ expression in \eqref{two_Col_eqn}. 
Combining these two statements yields the result of the~lemma.
\end{IEEEproof}

%%%%%%%%%%%%%%%%%%%%%%%%%%%%%%%%%%%%%%%%%%%%%%%%%%%%%%%%%%%%%%%%%%%%%%%%%%%%%%%%%%%%%%%%%%%%%%%%%%%%%%%%%%%%%%%%%%%%%%%%%%%
\subsection{Optimizing the Secondary User Delay}
In this subsection, we shift our attention to opportunistic spectrum
access in networks supporting real-time traffic, i.e., ORT traffic, 
%coined ORT, 
which have received attention only recently \cite{survey_Liang}. Towards this objective,
we investigate the problem of minimizing the average delay encountered by SU packets, $D_{s}$, subject to a constraint on the average PU packet delay, $D_{p}$.
%for a given arrival rate of the PU packets, $\lambda_p$,

We follow the same analysis presented in the previous subsection.
%Note that introducing a constraint on $D_p$ implies the stability condition of $Q_p$.
It is worth noting that the minimization of $D_s$ guarantees the stability of $Q_s$ unless the problem is infeasible.
Therefore, $Q_s$ and $Q_p$ stability conditions will be redundant and, hence, omitted from the problem formulation.
This step reduces the complexity of the optimization problem.
Consequently, the target constrained optimization problem is formulated as
\begin{eqnarray} \label{min_3}
    \displaystyle \textbf{P3: } \quad \min_{a, b} \!\!&&\!\! \frac{N_{s}}{\lambda_s} \nonumber\\
    \text{s.t. } \!\!&&\!\! 0 \leq a \leq 1 \nonumber\\
    \!\!&&\!\! 0 \leq b \leq 1 \nonumber\\
    \!\!&& \!\! \displaystyle \mu_p = h_{pd} + \left(1-h_{pd}\right) h_{ps} a \nonumber\\
    \!\!&& \!\! \displaystyle \frac{N_p+N_{sp}}{\lambda_p} \leq \psi,
\end{eqnarray}
where the objective function is the SU packet delay, $D_{s}$, and
$N_s$ is given by \eqref{eq:N_s2}.
\textbf{P3} is non-convex.
However, we can exploit the structure of \textbf{P3} to convert it to a set of quasiconvex optimization problems 
that can be solved for the optimal in an iterative manner as shown next.
%, in which the objective function is quasiconvex and the constraints are convex \cite{ConvxOpt}.
%
Following the same approach applied in the previous subsection, \textbf{P3} can be solved iteratively as follows
\begin{algorithmic}
\FOR{$\displaystyle\mu_p = h_{pd} : \delta : h_{pd} + \left(1 - h_{pd}\right)h_{ps}$}
\vspace{-3mm}
\STATE{
    \begin{eqnarray} \label{min_4}
    \displaystyle \textbf{P4: } \quad g_{4}\!\left(\mu_p\right)=\min_{b} \!\!&&\!\! \frac{N_{s}}{\lambda_s} \nonumber\\
    \text{s.t. } \!\!&&\!\! 0 \leq b \leq 1 \nonumber\\
    \!\!&& \!\! \displaystyle \phi_{\psi} \leq 0
\end{eqnarray}
} \ENDFOR \RETURN  $\displaystyle\max_{\mu_p} g_{4}\!\left(\mu_p\right)$.
\end{algorithmic}
Once again, we can see that this optimization problem is a low complexity line search in the interval 
$\left[\:\!h_{pd},\: h_{pd} \!+\! \left(1\!-\!h_{pd}\right)h_{ps}\:\!\right]$.
It can be shown through \eqref{eq:N_s2} that $D_s$ is quasiconvex in $b$.
%and substitutions from \eqref{eq:Ns_paramt2},
Since the objective function of \textbf{P4} is quasiconvex and the constraints are convex, 
\textbf{P4} is a quasiconvex optimization problem for each iteration on~$\mu_p$~\cite{ConvxOpt}, %, with a variable $b$.
and its solution is characterized by the following lemma.
\begin{lemma} 
\label{lemma1}
For a given $\mu_p$, the optimal solution of \textbf{P4}, $b^*\!\left(\mu_p\right)$, is equal to 
the the optimal solution of \textbf{P2}.
\end{lemma}
\begin{IEEEproof} 
\label{proof1}
In pursuance of solving \textbf{P4}, we delve into the 
relationship between the optimization problems \textbf{P2} and \textbf{P4}.
In the former problem, it is obvious that we maximize an objective 
function that monotonically increases with $b$. 
However, in the latter problem, we minimize an objective function 
that monotonically decreases with $b$. This can be readily 
verified by evaluating the first derivative of $D_{s}$ with respect to $b$. 
It can be shown that
\vspace{-1mm}
\begin{equation} 
\label{1st_derv}
\displaystyle\frac{\partial D_s}{\partial b} \!=\!
\frac
{- h_{sd} \: \mu_p \! \left(
\lambda_p \lambda_s^2 \left(1\!\!-\!\!\mu_p\right) \! \left(\mu_p\!\!-\!\!\lambda_p\right) 
\!+\! 
\left(\lambda_s\!\!-\!\!\lambda_s^2\right) \! \left(\mu_p\!\!-\!\!\lambda_p\right)^3\right)}
{\lambda_s \left(\lambda_s \mu_p\left(\mu_p\!\!-\!\!\lambda_p\right) 
\!-\! 
h_{sd} \left(\mu_p\!\!-\!\!\lambda_p\right)^2 b \right)^2},
\end{equation}
which is negative definite irrespective of the choice of~$b$ and, 
hence, $D_s$ monotonically decreases with $b$.
Taking into consideration that both problems have the same 
constraints, it can be asserted that \textbf{P2} and \textbf{P4}
are equivalent optimization problems; the feasible sets and the 
optimal solutions of both problems are identical. In other words, the problem of 
maximizing the SU throughput is equivalent to the problem of minimizing the SU 
packet delay for the adopted system model and cooperation policy.
%Thus, the optimal value $b^*\!\left(\mu_p\right)$ of \textbf{P2} and \textbf{P4} is identical.
This completes the proof of the lemma.
\end{IEEEproof}

%%%%%%%%%%%%%%%%%%%%%%%%%%%%%%%%%%%%%%%%%%%%%%%%%%%%%%%%%%%%%%%%%%%%%%%%%%%%%%%%%%%%%%%%%%%%%%%%%%%%%%%%%%%%%%%%%%%%%%%%%%%
%\section{Towards a Work-Conserving Cooperation Policy}

%%%%%%%%%%%%%%%%%%%%%%%%%%%%%%%%%%%%%%%%%%%%%%%%%%%%%%%%%%%%%%%%%%%%%%%%%%%%%%%%%%%%%%%%%%%%%%%%%%%%%%%%%%%%%%%%%%%%%%%%%%%
\section{Numerical Results}
%
%%\subsection{Non-WC Cooperation Policy}
In this section, we evaluate the performance of the proposed optimal 
%%non-WC 
cooperation policy for the cognitive radio network depicted in Fig.~\ref{fig:SysModel}.
We compare the proposed policy to a baseline cooperation policy (BL), coined 
``unconstrained partial cooperation policy"~\cite{Ashour_Journal}.
In~BL, the SU probabilistically cooperates with the PU in delivering its packets in $Q_{sp}$ with no constraint on the delay encountered by the PU packets. 
BL~can be formulated using the same objective functions of \textbf{P1} and \textbf{P3}, yet, subject only to queues stability constraints.
It should be emphasized that the BL yields better throughput and packet delay for the SU because it optimizes these performance metrics subject to the less stringent queues stability constraints. However, this, in turn, gives rise to poor PU performance in terms of arbitrarily large delays, as reported in \cite{Ashour_Journal}, since there is no delay constraint on the PU packets.
On the other hand, the proposed cooperation policy aims at optimizing the performance of the SU subject to a more stringent constraint, that is, the PU packet delay.
Therefore, unlike \cite{Ashour_Journal}, the proposed optimization problems balance the tradeoff between protecting the QoS of PUs and enhancing the QoS of SUs.

\begin{figure}
\centering
\includegraphics[width=1\linewidth]{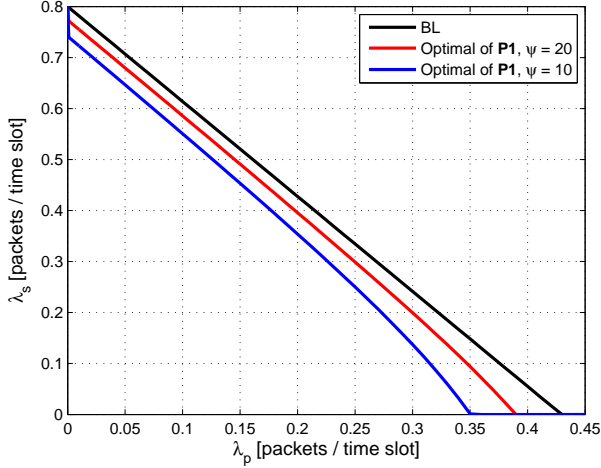}
\caption{The stable throughput region.}
\label{fig:fig2}
\vspace{-3mm}
\end{figure}

For the numerical results presented next, the following system parameters are used.
The successful packet reception probabilities between the nodes of the network are $h_{pd}=0.3$, $h_{ps}=0.4$, and $h_{sd}=0.8$.
Note that we set $h_{sd} > h_{pd}$ since the cooperation of the SU in delivering the PU packets does not make sense when $h_{sd} \leq h_{pd}$ and it would be better to transmit the PU packets to the destination through the direct link in such a case.
Furthermore, we solve \textbf{P1} and \textbf{P3} for different constraint values on the PU delay, $D_p \leq \psi \text{ where } \psi = 10 \text{ and } 20$.

In the first part of this section, we investigate the maximum stable throughput, i.e., the boundary of the stable throughput region, for the proposed optimal 
%%non-WC 
cooperation policy.
In~Fig.~\ref{fig:fig2}, we plot the stable throughput region of the system for different constraint values on the average delay experienced by the PU packets, $D_p$.
%In~Fig.~\ref{fig:fig2}, we plot the stable throughput region of the system, i.e., the maximum attainable service rate of the SU packets, $\mu_s$, versus the arrival rate of the PU packets, $\lambda_p$, for different constraint values on the average delay experienced by the PU packets, $D_p$.
%
Under the BL scheme, the SU enjoys higher throughput since there is no constraint on $D_p$. However, the PU experiences huge packet delay, i.e., $D_p \rightarrow \infty$.
It is worth mentioning that the violation of the QoS requirements of PUs (spectrum owners) is a serious problem in cognitive radio networks.
On the contrary, in the proposed optimal cooperation policy, when the PU delay constraint is introduced in \textbf{P1}, e.g, $D_p \leq 20$, the average packet delay experienced by the PU is guaranteed not to exceed $\psi = 20$.
Moreover, the system does not lose much in terms of the stable throughput region. 
Hence, the proposed problem formulation optimizes the SU throughput, but, at the same time, maintains a certain level of QoS for the PU.
However, protecting the PU QoS comes at the expense of a decrease in the SU stable throughput compared to the BL.
Furthermore, when the constraint on $D_p$ in~\textbf{P1} becomes tighter, i.e., $\psi$ decreases, the stable throughput region shrinks in order to satisfy this constraint.

\begin{figure}
\centering
\includegraphics[width=1\linewidth]{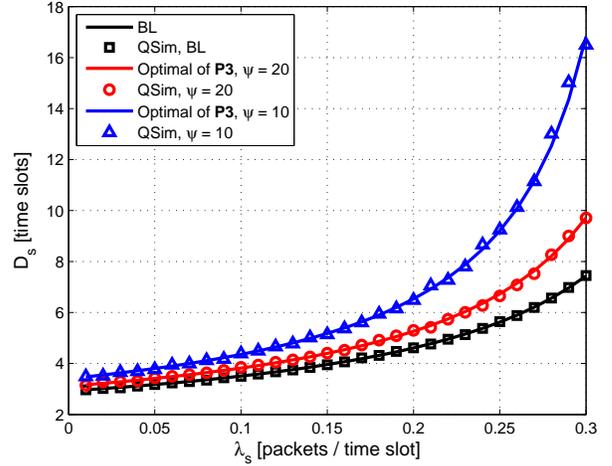}
\caption{The delay-throughput tradeoff at the SU, $\lambda_p=0.2$.}
\label{fig:fig3}
\vspace{-3mm}
\end{figure}

\begin{figure}
\centering
\includegraphics[width=1\linewidth]{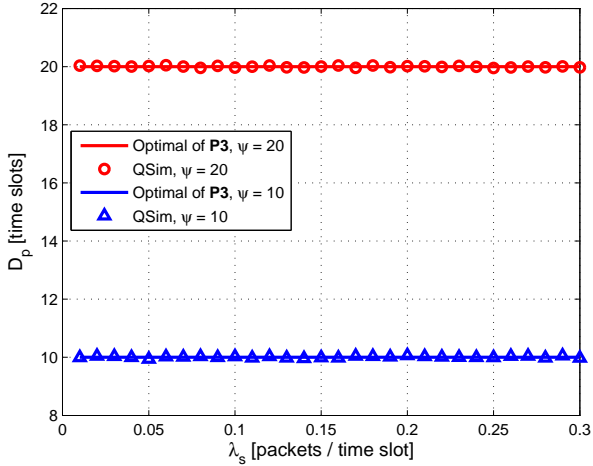}
\caption{The average PU packet delay versus the arrival rate of the SU packets, $\lambda_p=0.2$.
The corresponding PU delay of  the BL takes arbitrarily large values \cite{Ashour_Journal} and, hence, 
it is not plotted.}
\label{fig:fig4}
\vspace{-3mm}
\end{figure}

\begin{figure}
\centering
\includegraphics[width=1\linewidth]{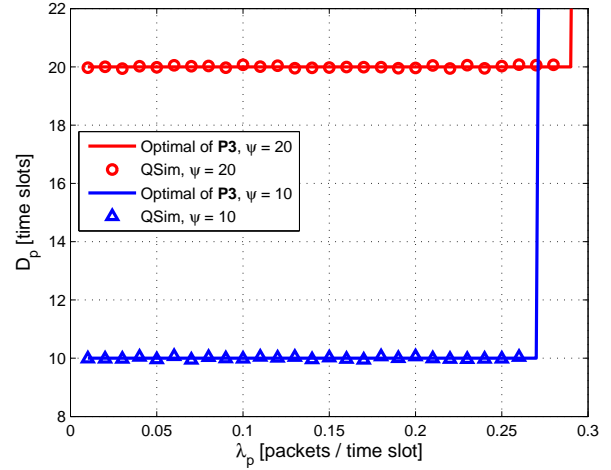}
\caption{The delay-throughput tradeoff at the PU, $\lambda_s=0.2$.
The corresponding PU delay of the BL takes arbitrarily large values \cite{Ashour_Journal} and, hence, 
it is not plotted.}
\label{fig:fig5}
\vspace{-3mm}
\end{figure}

Next, we shift our attention to the delay performance of the PUs and SUs for the proposed optimal 
%%non-WC 
cooperation policy.
We set $\lambda_p=0.2$ in~\cref{fig:fig3,fig:fig4}, and $\lambda_s=0.2$ in~\cref{fig:fig5}.
The average delays are computed via the optimal solution of~\textbf{P3} and the queue simulation (QSim).
The packet delays are averaged over $10^5$ time slots in the QSim.
The results of the optimal solution of~\textbf{P3} coincide with those obtained from the QSim as shown in~\cref{fig:fig3,fig:fig4,fig:fig5}.
%W%\cref{fig:fig3,fig:fig4,fig:fig5}

%%%Fig.~\ref{fig:fig3} depicts the relationship between the average delay experienced by the SU packets, $D_s$, versus the arrival rate of the SU packets, $\lambda_s$, 
Fig.~\ref{fig:fig3} depicts the delay-throughput tradeoff at the SU
for different constraint values on the average delay experienced by the PU packets, $D_p$.
It is obvious that $D_s$ monotonically increases with $\lambda_s$ for all cooperation policies.
Furthermore, we can see that the SU delay of the BL is lower than the SU delay introduced by the proposed optimal cooperation policy.
However, the corresponding PU packet delay of the BL takes arbitrarily large values, i.e., $D_p \rightarrow \infty$. 
Unlike BL, the proposed optimal cooperation policy minimizes the SU delay and guarantees that the PU packet delay remains bounded, i.e., less than or equal to $\psi$, as shown in Fig.~\ref{fig:fig4}. 
In other words, assuring a certain level of QoS for the PU while enhancing the SU QoS comes at the expense of an increase in the SU delay compared to the BL.
We can also see from Fig.~\ref{fig:fig3} that when the constraint on $D_p$ becomes tighter, i.e., $\psi$ decreases, the resulting value of the objective function of \textbf{P3}, $D_s$, increases. 
The reason for this behavior lies behind the strict constraint on the PU delay that forces the system to choose $Q_{sp}$ more often and, hence, a lower probability of choosing $Q_s$ is obtained from the solution of \textbf{P3}, i.e., lower $b$, in order to satisfy the constraint. Therefore, $D_s$ increases.

Fig.~\ref{fig:fig4} shows the average delay of the PU packets, $D_p$, versus the arrival rate of the SU packets, $\lambda_s$, for different constraint values on $D_p$. 
It should be noted that the PU delay of the BL takes arbitrarily large values, i.e., $D_p \rightarrow \infty$,
since there is no delay constraint on the PU packets in the BL.
%Note that the PU delay of the BL is not plotted since the delay in this case takes arbitrarily large values.
In the proposed optimal cooperation policy, on the other hand, it is evident that the PU delay constraint is always satisfied with equality,  i.e., $D_p = \psi$. In other words, the constraint on $D_p$ is satisfied at the boundary of the feasible set of \textbf{P3} in order to reach the minimum value of the objective function, i.e., the minimum SU delay.

Fig.~\ref{fig:fig5} captures the delay-throughput tradeoff at the PU for different constraint values on the average delay encountered by the PU packets, $D_p$.
%
%%The proposed optimal cooperation policy always satisfies the PU delay constraint with equality, i.e., $D_p = \psi$.
%%In other words, the constraint on $D_p$ is satisfied at the boundary of the feasible set of \textbf{P3} in order to reach %%the minimum value of the objective function, i.e., the minimum SU delay.
%
%Recall that the PU delay corresponding to the BL takes arbitrarily large values, i.e., $D_p \rightarrow \infty$,
%since there is no delay constraint on the PU packets in the BL.
%
%Therefore, the proposed cooperation policy is practical in the sense that it optimizes the SU QoS without violating the constraint on the PU QoS.
%
We can see that there is a maximum value for $\lambda_p$ after which \textbf{P3} becomes infeasible, e.g., the sudden jump of $D_p$ at $\lambda_p = 0.29$ when $\psi = 20$, and at $\lambda_p = 0.27$ when $\psi = 10$.
In other words, there are no values for $a$ and $b$ that stabilize the queues of the system under the PU delay constraint.
Furthermore, when the constraint on $D_p$ becomes tighter, the value of $\lambda_p$ at which the system reaches the unstable state becomes smaller. 
It is worth noting that these values of $\lambda_p$ for the different constraint values on $D_p$ are consistent with the stable throughput region plotted in Fig.~\ref{fig:fig2} when $\lambda_s=0.2$.

%%%%%%%%%%%%%%%%%%%%%%%%%%%%%%%%%%%%%%%%%%%%%%%%%%%%%%%%%%%%%%%%%%%%%%%%%%%%%%%%%%%%%%%%%%%%%%%%%%%%%%%%%%%%%%%%%%%%%%%%%%%
%%\subsection{Suboptimal WC Policy $\&$ WC-Policy-based Algorithm}

%%%%%%%%%%%%%%%%%%%%%%%%%%%%%%%%%%%%%%%%%%%%%%%%%%%%%%%%%%%%%%%%%%%%%%%%%%%%%%%%%%%%%%%%%%%%%%%%%%%%%%%%%%%%%%%%%%%%%%%%%%%
\section{Conclusion}
In this paper, we studied cooperative cognitive radio
networks with the objective 
of optimizing the QoS of SUs that support opportunistic real-time traffic while sustaining the typical QoS of PUs.
%%for both non-work-conserving (non-WC) and work-conserving (WC) cooperation policies.
%%In the first part of the paper,
We formulated two optimization problems subject to a common
constraint on the maximum packet delay that the PU can tolerate
for a non-work-conserving (non-WC) cooperation policy, motivated by its mathematical tractability.
The objective of the first problem is to maximize the SU throughput, 
whereas the objective of the second problem is to minimize 
the SU average packet delay.
We proved that both optimization problems are 
equivalent with the same feasible sets and optimal solutions. 
%
%The simulation results have shown that the proposed cooperation 
%scheme enables the tradeoff between protecting the PU QoS and 
%enhancing the SU QoS.
%
The numerical results demonstrated that the 
stable throughput region of the proposed optimal 
cooperation policy approaches that 
of the unconstrained partial cooperation policy, depending on 
the constraint imposed on the PU packet delay. 
Furthermore, the average PU packet delay of the proposed optimal
cooperation policy is much lower than the one of the unconstrained partial 
cooperation policy. However, this comes at the expense of an
increase in the average SU packet delay and the amount of increase
depends on how tight the constraint on the PU packet delay is.

%%In the second part of the paper, we extended our model to study a WC
%%cooperation policy and investigated the problem of optimizing the
%%SU packet delay subject to a constraint on the PU packet delay.
%%Due to the sheer complexity of deriving closed-form expressions for
%%the average delay of the PU and SU packets for a WC
%%policy, due to multiple queues interaction, we reformulate the target optimization problem whereby the
%%sum of the average length of the SU data queue and SU relaying queue
%%is minimized subject to a constraint on the PU packet delay
%%of the non-WC policy. As a result, an
%%upper bound on the optimal SU delay is obtained.
%%Through our analysis as well as our simulation results, we
%%show that the proposed suboptimal WC policy outperforms the
%%non-WC policy proposed in the first part of the paper. Furthermore, we
%%propose a novel theoretically-founded WC-policy-based algorithm
%%in pursuance of approaching the optimal SU delay of 
%%the target optimization problem.

%W%Future work includes studying the problem of optimizing the SU performance 
%W%with PU QoS provisioning under imperfect spectrum sensing setting.
%W%Furthermore, the research on developing new mathematical tools for characterizing 
%W%delays for three interacting queues is still open for more investigation.

%%%%%%%%%%%%%%%%%%%%%%%%%%%%%%%%%%%%%%%%%%%%%%%%%%%%%%%%%%%%%%%%%%%%%%%%%%%%%%%%%%%%%%%%%%%%%%%%%%%%%%%%%%%%%%%%%%%%%%%%%%%
%%\appendix

%%%%%%%%%%%%%%%%%%%%%%%%%%%%%%%%%%%%%%%%%%%%%%%%%%%%%%%%%%%%%%%%%%%%%%%%%%%%%%%%%%%%%%%%%%%%%%%%%%%%%%%%%%%%%%%%%%%%%%%%%%%
\begingroup
\newif\ifgobblecomma
\gobblecommafalse %default
\edef\FZ{?}
\edef\KM{,}
\catcode`?=\active
\catcode`,=\active
\def?{\FZ\gobblecommatrue}
\def,{\ifgobblecomma\gobblecommafalse\else\KM\fi}
\bibliographystyle{IEEEbib}
\linespread{1}
\bibliography{myRef}
\endgroup
\end{document}